\documentclass[aps,prc,groupedaddress,twocolumn,showpacs,amsmath,amssymb]{revtex4}
\usepackage{graphics}% Include figure files
\usepackage{dcolumn}% Align table columns on decimal point 
\usepackage{longtable}
\begin{document}
\title{Role of the direct processes in low-energy deuteron interactions}

%%%%%\author{\firstname{Marilena} \lastname{Avrigeanu}\inst{1}\fnsep\thanks{\email{marilena.avrigeanu@nipne.ro}}% \and
%%%%%       \firstname{Vlad} \lastname{Avrigeanu}\inst{1}\fnsep\thanks{\email{vlad.avrigeanu@nipne.ro}}\and
%%%%%			\firstname{Cristian} \lastname{Costache}\inst{1}\fnsep\thanks{\email{cristian.costache@nipne.ro}}
%%%%%}
\author{M.~Avrigeanu} \email{marilena.avrigeanu@nipne.ro}
\author{V.~Avrigeanu}
\author{C.~Costache}

%%%%%\institute{Horia Hulubei National Institute for Physics and Nuclear Engineering, P.O. Box MG-6, 077125 Bucharest-Magurele, Romania}
\affiliation{Horia Hulubei National Institute for Physics and Nuclear Engineering, P.O. Box MG-6, 077125 Bucharest-Magurele, Romania}

%%%%%\abstract{
\begin{abstract}
An extended analysis of the key role of direct interactions, i.e., breakup, stripping and pick-up processes, has been carried out for deuteron-induced reactions. Particular comments concern the deuteron breakup which is the dominant mechanism involved in surrogate reactions on heavy nuclei, around the Coulomb barrier. 
%%%%%}
\end{abstract}

\maketitle
\section{Introduction}
\label{intro}

At present increased deuteron-data needs follow the demands of on-going strategic research programs (ITER, IFMIF, SPIRAL2-NFS \cite{iter}) that involve deuteron beams, while the corresponding experimental and evaluated data are less extensive and accurate than for neutrons. %Moreover, the analysis of low-energy deuteron reactions in terms of usual reaction models is challenging due to deuteron breakup (BU) as a result of its weak binding energy. 
There are currently many efforts to improve the description of deuteron reactions (e.g., \cite{thBU}) also due to the use of $(d,p\gamma)$, and $(d,pf)$ surrogate reactions for neutron capture $(n,\gamma)$ and  induced fission $(n,f)$ cross sections investigations respectively, of interest for breeder reactors. 

The consistent account of the direct interactions (DI) role in deuteron interaction process \cite{AlCudpap,dpap2}, i.e., the breakup (BU) and direct reactions (DR) (stripping and pick-up), forms the object of this work for nuclei from $^{27}$Al to $^{238}$U. While the above-mentioned references include a detailed description of the involved model assumptions and consistent parameters sets that were either established or validated using various independent data, we briefly mention here only the updated main points of these analyzes. 

\section{Deuteron breakup}
\label{sec-1}

Concerning the physical picture of the deuteron breakup in the Coulomb and nuclear fields of the target nucleus, two distinct processes are considered in this respect, namely the elastic breakup (EB) in which the target nucleus remains in its ground state and none of the deuteron constituents interact with it, and the inelastic breakup or breakup fusion (BF), where one of these deuteron constituents interacts nonelastically with the target nucleus. The reactions induced by the BF nucleons lead to different compound nuclei than following the incoming deuterons, the partition of the BF cross section among various residual nuclei being triggered by the energy spectra of the BF nucleons and the excitation functions of the pre-equilibrium (PE) and compound-nucleus (CN) reactions induced by these nucleons on the target nuclei \cite{AlCudpap,dpap2}. %Also, the excitation energies of the CNs which are formed by the BF nucleons and the incident deuteron, respectively, are quite different \cite{AlCudpap,dpap2}. 

\begin{figure*}
%%%%%\centering
%%%%%\includegraphics[width=16.9cm]{Fig_BFE_E}
\resizebox{2.06\columnwidth}{!}{\includegraphics{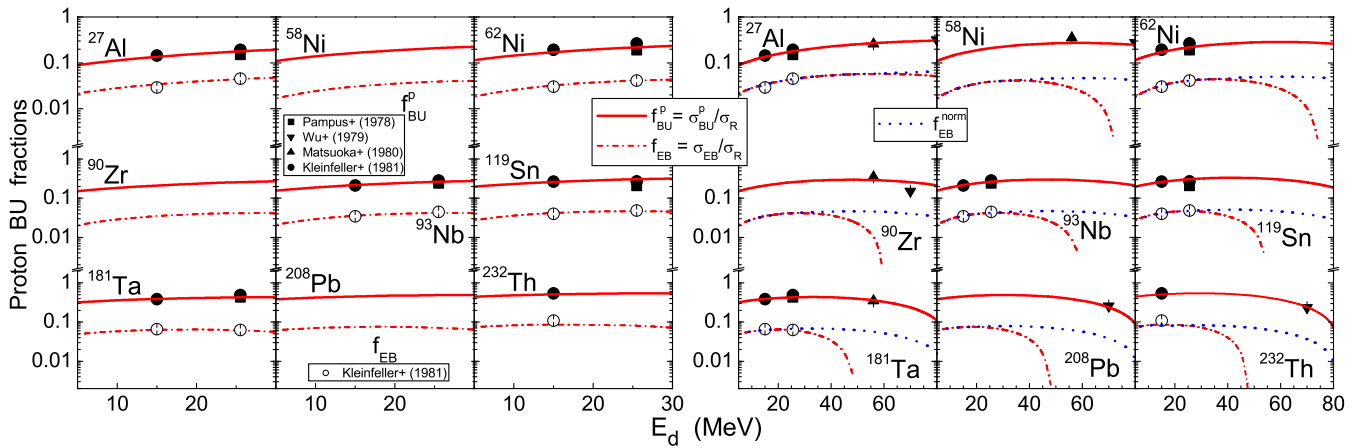}}
\caption{Comparison of experimental \cite{pamp78} breakup proton-emission (solid symbols) and elastic-breakup (open circles) fractions, and the corresponding parametrizations \cite{breakup} (solid and dash-dotted curves, respectively), for deuterons on nuclei from $^{27}$Al to $^{232}$Th at incident energies up to 30 MeV (left) and 80 MeV (right), as well as the normalized EB fractions (dotted) for the higher energies (right).}
\label{Fig_BFE_E}
\end{figure*}

The BU cross sections have been obtained from an empirical parametrization \cite{breakup} of both the total (EB+BF) breakup $proton$-emission and EB fractions,  $f_{BU}^{p}$ = $\sigma^p_{BU}/\sigma_R$ and $f_{EB}$=$\sigma_{EB}/\sigma_R$, respectively, where $\sigma_{R}$ is the deuteron total-reaction cross section. At the same time the inelastic-breakup fraction is given by the difference $f_{BF}^{p}$=$f_{BU}^{p}$-$f_{EB}$.
The comparison of the parametrization values and experimental $f_{BU}^{p}$ and $f_{EB}$ fractions  \cite{pamp78}, for target nuclei from $^{27}$Al to $^{232}$Th, is shown in Fig.~\ref{Fig_BFE_E} for deuteron energies up to 30 MeV (left) corresponding to the experimental data basis \cite{pamp78} for the $f_{EB}$ parametrization, as well as in the extended energy range up to 80 MeV (right) of the experimental data basis for the $f_{BU}^{p}$ parametrization.

A first comment concerns the apparent decrease of $f_{EB}$ at incident energies beyond the range within which it was established. Since this trend has  been opposite to that of $f_{BU}^{p}$ fraction, in the absence of any available deuteron EB data at $E_d$$>$30 MeV, the correctness of the extrapolation in the case of the EB parametrization has been checked by the comparison of its predictions and results of the Continuum-Discretized Coupled-Channels (CDCC) method \cite{maCDCC}. %Two different features have been thus pointed out. A good  agreement was obtained between the CDCC results, the EB parametrization predictions \cite{breakup}, and the data systematics \cite{pamp78} at energies where these data were available, (e.g., Figs. 2 and 3 of Ref. \cite{maCDCC}), which validates the empirical parametrization. 
The good  agreement between the CDCC results, EB parametrization predictions \cite{breakup}, and data systematics \cite{pamp78} has increased the confidence on both CDCC approach and empirical-parametrization basic grounds at energies of the available data (Figs. 2 and 3 of Ref. \cite{maCDCC}).

%However
On the other hand, the behavior of CDCC results over 30 MeV has been in agreement only with the $f_{BU}^{p}$ parametrization. Thus it has resulted that the extension of the $f_{EB}$ empirical parametrization beyond the energies at which the corresponding data were available should be considered with caution. Therefore, in order to avoid laborious CDCC calculations for each target/energy of interest, we have adopted a normalized EB fraction for the energies beyond its maximum by taking into account the CDCC results and the behavior of the $f_{BU}^{p}$ fraction. Hence, we have chosen to keep unchanged the  ratio of the EB and BU fractions for the incident energies above the maximum of EB fraction $f_{EB}(E_{d,max})$ by means of the relation
\begin{equation}\label{eq:1}
      f_{EB}^{norm}(E_d) = f_{BU}^{p}(E_d) \frac{f_{EB}(E_{d,max})}{f_{BU}^{p}(E_{d,max})} , \:\:\:\: E_d > E_{d,max} \:\:,
\end{equation}
\noindent
where the energy $E_{d,max}$ corresponds to the above-mentioned maximum. Thus, the normalized EB fraction follows the behavior of total  breakup proton-emission fraction as it is also shown in Fig.~\ref{Fig_BFE_E}, in agreement with the CDCC calculation results \cite{maCDCC}.  This $f_{EB}$ normalization is of particular interest, at deuteron energies above $\sim$50 MeV and especially for heavier target nuclei where $E_{d,max}$ is lower, for both inelastic- and total-breakup fractions \cite{breakup}: 
\begin{equation}\label{eq:2}
 f_{BF}^{p/n}  =  f_{BU}^{p/n}  - f_{EB}^{norm} ,\:\: \:\:    f_{BU}  =  2 f_{BU}^{p/n} - f_{EB}^{norm} , 
\end{equation}
assuming equal proton and neutron BU cross sections.

On the other hand, the  experimental systematics of $f_{BU}^{p}$ \cite{pamp78} includes only one data for heavy nuclei ($A$$>$200) at incident energies around the Coulomb barrier, which are of great interest for surrogate reaction study, namely for $^{232}$Th at $E_d$=15 MeV. The present parametrization
describes well this data as well as the one at a medium energy of 70 MeV (Fig~\ref{Fig_BFE_E}). However, within the energy range between $\sim$25 and 41 MeV, i.e. around the $E_{d,max}$$=$32 MeV  for this target nucleus, the corresponding total BU fraction exceeds unity (e.g., $f_{BU}$=1.0215 at $E_{d,max}$). Consequently, we have adopted for $A$$>$200 the additional constraint that the $f_{BU}$ fraction should be less than 90\%. It has been included as well as the normalization by the use of Eq. ($\ref{eq:1}$) in the code TALYS$-$1.8 \cite{talys}, corresponding to the value $\bf{2}$ of the $breakupmodel$ option for the BU model calculations.

\section{Direct reactions}
\label{sec-2}

Usually neglected or very poorly taken into account, the direct reactions play an important role in the deuteron interactions at energies around the Coulomb barrier. Their contribution is important for the first--chance emitted particle cross section \cite{AlCudpap,dpap2,breakup}. 

\begin{figure*}
\resizebox{2.06\columnwidth}{!}{\includegraphics{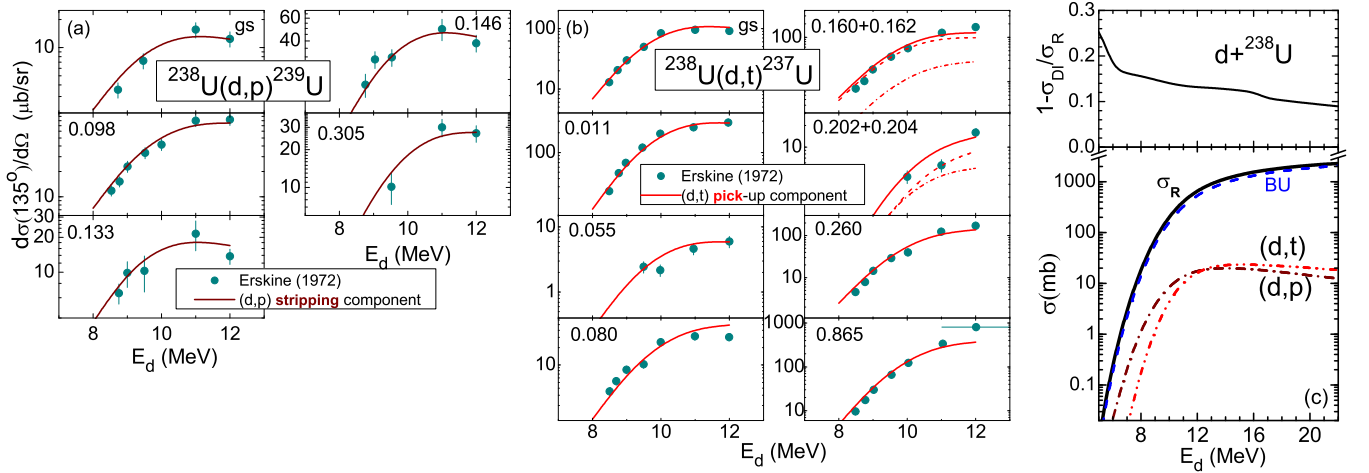}}
\caption{Comparison of calculated (solid curves) and measured \cite{erskine} excitation functions for the populations at 135$^{o}$ of low-lying levels in (a) $^{239}$U and (b) $^{237}$U, through $(d,p)$ and $(d,t)$ direct reactions, respectively. (c) Total-reaction (solid curves), breakup (dashed curves), stripping $(d,p)$ (dash-dotted curves), and pick-up $(d,t)$ (dash-dot-dotted curves) cross sections for deuterons on $^{238}$U (bottom), and the reduction factor of the deuteron flux going towards statistical processes (bottom).}
\label{Fig_FR-238Ud}       
\end{figure*}

The $(d,p)$ and $(d,n)$ stripping, and $(d,t)$ and $(d,\alpha)$ pick-up cross sections were calculated using the distorted-wave Born approximation (DWBA) method, within FRESCO code \cite{FRESCO}, with details given elsewhere \cite{AlCudpap,dpap2,breakup}. Their correctness has been validated by the suitable description of the experimental specific data, e.g. angular distributions and double-differential cross sections.
Therefore, the description of the measured \cite{erskine} populations of low-lying levels in $^{239}$U and $^{237}$U through $^{238}$U(d,p)$^{239}$U and $^{238}$U(d,t)$^{237}$U stripping and pick-up reactions, respectively, shown in Fig.~\ref{Fig_FR-238Ud}(a,b),  validates the correctness of the corresponding $(d,p)$ and $(d,t)$ total excitation functions in Fig.~\ref{Fig_FR-238Ud}(c, bottom). Sum of the total $(d,p)$, $(d,t)$, and BU cross sections gives a lower limit of the DI contribution to the deuteron interaction with $^{238}$U target nucleus. 

Finally, the deuteron total-reaction cross section that remains to be available for the PE+CN lengthy mechanisms has to be corrected for the incident-flux leakage through DI processes, {\i.e.} the breakup, stripping and pick-up, by a reduction factor \cite{AlCudpap,dpap2,breakup}:
\begin{equation}\label{eq:3}
       1 - \frac{\sigma_{BU} + \sigma_{(d,p)} + \sigma_{(d,t)}}{\sigma_R} = 1 - \frac{\sigma_{DI}}{\sigma_R}. 
\end{equation}
\noindent
The reduction factor shown on top of Fig.~\ref{Fig_FR-238Ud}(c) points out the DI dominant role (mainly breakup) in the deuteron interaction with $^{238}$U around Coulomb barrier, which is a specific feature in the case of heavy nuclei \cite{Pad,varna}.

\section{DI effects on surrogate reactions}
\label{sec-3}

Since last decade the surrogate reaction method \cite{escher,wilson,ducasse} has been intensively utilized to measure mainly $(n,\gamma)$ and $(n,f)$ cross sections by means of the surrogate reactions like $(d,p\gamma)$ and $(d,pf)$. 
The "desired" $(n,\gamma)$ cross section for a target nucleus $A$ is given in terms of the CN formation cross section $\sigma^{CN}_{n}(E_{ex},J,\pi)$ and the branching ratio $G^{CN}_{\gamma}(E_{ex},J,\pi)$ toward the desired $\gamma$ outgoing channel:
\begin{equation}\label{eq:4}
\sigma_{n,\gamma}(E_{n}) = \Sigma_{J,\pi}\: \sigma^{CN}_{n}(E_{ex},J,\pi) G^{CN}_{\gamma}(E_{ex},J,\pi),
\end{equation}
where $J,\pi$ are the spin and parity of the excited state $E_{ex}$.% of the decaying CN.

Usually $\sigma^{CN}_{n}(E_{ex},J,\pi)$ is provided by a neutron optical model potential, while 
theoretical branching ratios
$G^{CN}_{\gamma}(E_{ex},J,\pi)$ 
%requires accurate information on the Hauser-Feshbach model ingredients. 
are often quite uncertain \cite{escher}. They are extracted in the surrogate method 
%Attempts to avoid related difficulties tried to do so 
by measuring in $(d,p\gamma)$ surrogate reaction the probability $P_{d,p\gamma}^{exp}(E_{ex})$ for the formation of the same excited-nucleus states, with the same specific $E_{ex}$ and $J$, $\pi$ values, decaying through the $\gamma$-emission channel.
The probability of this surrogate excited-nucleus decay through the $\gamma$ channel is \cite{escher}:
\begin{equation}\label{eq:5}
P_{d,p\gamma}(E_{ex}) =  \Sigma_{J,\pi}  \: F^{CN}_{d,p}(E_{ex},J,\pi) G^{CN}_{\gamma}(E_{ex},J,\pi) , 
\end{equation}
where $F^{CN}_{d,p}(E_{ex},J,\pi)$ is the corresponding probability for the formation of desired excited nucleus in the surrogate reaction.
Moreover, $P_{d,p\gamma}(E_{ex})$ is obtained experimentally by measuring the total number of the surrogate events given by, e.g., proton spectrum, and the number of coincidences between the surrogate ejectile and the $\gamma$-decay channel, namely, the number of the $p$-$\gamma$ coincidences:
\begin{equation}\label{eq:6}
P^{exp}_{d,p\gamma}(E_{ex}) = \frac{N^{coincidences}_{p,\gamma}(E_{ex})}{N^{surrogate events}_{d,p}(E_{ex})}.
\end{equation}
However, it is obvious that $P^{exp}_{d,p\gamma}$ of Eq. ($\ref{eq:5}$) {\it does not included the contributions of CN mechanism only}. 
Finally, simplifying the theoretical frame by additional assumptions of (i) similar $J^{\pi}$ distributions in both desired and surrogate reactions, and (ii) no $J^{\pi}$-dependence of the decay probabilities $G_{\gamma}(E_{ex},J,\pi)$ (the Weisskopf-Ewing model), the desired neutron capture cross section becomes \cite{escher}:
\begin{equation}\label{eq:7}
\sigma_{n,\gamma}(E_{n}) = \sigma^{CN}_{n}(E_{n})P^{exp}_{d,p\gamma}(E_{ex}).
\end{equation}
The validation test of the deuteron surrogate method by comparing well known $(n,\gamma)$ reaction cross sections with those obtained by analysis of surrogate $(d,p\gamma)$ reaction stressed out large discrepancies (\cite{wilson,ducasse} and Refs. therein) which rise question marks on this method. 
The reported results at variance from the surrogate $^{238}$U$(d,p\gamma)$$^{239}$U reaction at the incident energy of 15 MeV, and 'desired' $^{238}$U$(n,\gamma)$$^{239}$U reaction, for incident-neutron energies from 0 to 1.5 MeV, by Ducasse {\it et al.} \cite{ducasse} are involved hereafter particularly to underline the DI effects.
\begin{figure} [b]
%%%%%\centering
%%%%%\includegraphics[width=8.cm]{Fig_3a_rev}
\resizebox{1.00\columnwidth}{!}{\includegraphics{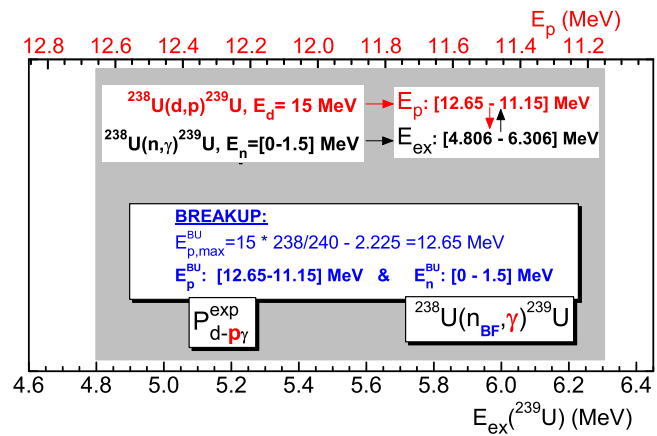}}
\caption{Correlation of the residual excitation energies in $^{238}$U$(n,\gamma)$$^{239}$U desired reaction at $E_n$=0--1.5 MeV \cite{ducasse} (bottom, in black) and energies of emitted protons in $^{238}$U$(d,p\gamma)$$^{239}$U surrogate reaction at $E_d$=15 MeV (top, in red), as well as the breakup-nucleon energies for 15 MeV deuterons (insert, in blue) and the perturbed cross-section measurements.}
\label{Fig_3a}     
\end{figure}

First, the decay probabilities $P_{d,p\gamma}(E_{ex})$ of the residual nucleus $^{239}$U were measured by Ducasse {\it et al.} at the excitation energies between the corresponding neutron-binding energy $S_n$=4.806 MeV and 1.5 MeV above it. The protons from 15 MeV deuteron-induced $(d,p\gamma)$ reaction on the $^{238}$U target nucleus, corresponding to this excitation energy range, have energies between 12.65 and 11.15 MeV (Fig.~\ref{Fig_3a}). These energies are matched also by the 15 MeV deuteron-breakup protons, with the maximum value $E_{p,max}^{BU}$=12.65 MeV. Thus, these BU protons may affect the the $P_{d,p\gamma}^{exp}(E_{ex})$ measurement.  

Second, the BU protons are correlated with the BU neutrons with energies between 0 and 1.5 MeV. Moreover, the BF component of these BU neutrons may interact with the $^{238}$U target nucleus, populating actually the desired  compound nucleus $^{239}$U at the same excitation energies of interest. Thus, the $\gamma$-decay of the $^{239}$U$^*$ populated through $^{238}$U(n$_{BF}$,$\gamma$) reaction may also affect, together with the companion BF protons, the measured $p$-$\gamma$ coincidence events and, finally, the $P_{d,p\gamma}^{exp}(E_{ex})$ measurement.  

Third, the probability $F^{CN}_{d,p}(E_{ex},J,\pi)$, given by Eq. ($\ref{eq:5}$) for forming the excited nucleus $^{239}$U, is decreased due to the incident flux leakage through DI, according to Eq. ($\ref{eq:3}$) and Fig~\ref{Fig_FR-238Ud}(c), as well as the PE processes which may precede the population of the excited nucleus $^{239}$U through the CN reaction mechanism d+$^{238}$U$\to$$^{240}$Np$^{*}$$\to$p+$^{239}$U$^*$.

Actually, it is a laborious task to select the contribution of the CN mechanism, i.e., {\it the only one considered in the theoretical frame of surrogate reactions}, from the measured probabilities $P_{d,p\gamma}^{exp}(E_{ex})$. It involves several corrections concerning the processes left out.
In this respect, the simple correction applied in Ref. \cite{ducasse} only to the single proton spectrum in Eq. ($\ref{eq:6}$) does not take into account neither the strongly reduced $F^{CN}_{d,p}(E_{ex},J,\pi)$ probability of forming $^{239}$U$^*$ nucleus by the deuteron flux leakage through DI and PE processes, nor the contributions of other than CN mechanism to the population of this nucleus. 

Moreover, the assumption concerning the equality of the branching ratios for the deuteron surrogate and the neutron-induced reactions does not hold, due to the population and decay differences between the excited and compound nuclei formed in surrogate and desired reactions, respectively \cite{chiba}. However, one should be more careful in assuming that the failure of the surrogate-method validation tests follows the use of the too weak Weisskopf-Ewing approximation \cite{wilson,ducasse}. It is also even the use of the Hauser-Feshbach formalism alone, within deuteron-surrogate reactions analysis, which can not lead to the expected good results in the absence of the unitary account of all BU+DR+PE+CN reaction mechanisms \cite{varna}. 

Finally it is obvious that the hard approximations that led to Eq. ($\ref{eq:7}$) are not appropriate for deuteron induced reactions. Therefore, the theoretical frame of deuteron surrogate method should be revised for a consistent account of the reaction mechanisms involved in the complex deuteron interactions.

\section{Conclusions}
\label{sec-4} 

The present work discussed a deeper analysis of the key role of DI, particularly of the breakup mechanism, in deuteron-induced reactions. 
Firstly, a normalization of the parametrized \cite{breakup} EB fraction for $E_d$ values beyond the energy range within it was established, $E_d$$\approx$25-30 MeV, has been provided in order to follow the behavior of the total BU proton fraction, in agreement with CDCC results. 

Next we have analyzed the contributions of the breakup, stripping, and pick-up reaction mechanisms to d+$^{238}$U interaction process, at energies around the Coulomb barrier, to illustrate the importance of direct interaction processes.

Finally, the validation test of the $^{238}$U(d,p$\gamma$) surrogate reaction has been discussed. The disregard of the DI mechanisms in the theoretical frame of deuteron surrogate-reaction method should be considered the main reason for the failure of the $(d,p\gamma)$ validation tests  \cite{wilson,ducasse}. However, the surrogate ratio method should be valid in the presence of the deuteron breakup without assuming its specific effects, due to their possible cancellations \cite{chiba}.%, while "it seems to be hopeless to apply the surrogate method for the capture cross-section measurements" \cite{chiba}.

This work was partly supported by Fusion for Energy (F4E-GRT-168-02) and Unitatea Executiva pentru Finantarea Invatamantului Superior, a Cercetarii, Dezvoltarii si Inovarii (Project No. PN-II-ID-PCE-2011-3-0450).

\end{document}